\documentstyle[aps,epsfig,twocolumn]{revtex}
\input epsf.sty

\title{Quantum Computers and Quantum Coherence}

\author{David P. DiVincenzo}
\address{
IBM Research Division,
Thomas J. Watson Research Center,
P. O. Box 218,
Yorktown Heights, NY 10598, USA
}
\author{Daniel Loss}
\address{
Department of Physics and Astronomy,
University of Basel,
Klingelbergstrasse 82,
4056 Basel, Switzerland
}

\date{\today}

\begin{document}
\twocolumn[\hsize\textwidth\columnwidth\hsize\csname
@twocolumnfalse\endcsname

\maketitle

\begin{abstract}
If the states of spins in solids can be created, manipulated, and
measured at the single-quantum level, an entirely new form of
information processing, quantum computing, will be possible.  We first
give an overview of quantum information processing, showing that the
famous Shor speedup of integer factoring is just one of a host of
important applications for qubits, including cryptography, counterfeit
protection, channel capacity enhancement, distributed computing, and
others.  We review our proposed spin-quantum dot architecture for a
quantum computer, and we indicate a variety of first generation
materials, optical, and electrical measurements which should be
considered.  We analyze the efficiency of a two-dot device as a
transmitter of quantum information via the  propagation of
qubit carriers (i.e. electrons) in a Fermi sea.
\end{abstract}

\vskip2pc]
\narrowtext

\section{Information processing and quantum mechanics}

While we will spend much of this chapter considering fairly
specifically the application of quantum magnetic systems to quantum
computing, we want to first review more broadly the potential
``quantum revolution'' that is brewing in the area of information
science.  It is amusing for a physicist to note that quantum mechanics
is now being taught as part of the standard curriculum in a growing
number of graduate computer science programs!  Why would computer
scientists find it necessary to take up such an esoteric study from a
different field?  The problems which have interested them have nothing
to do with the quantum world, and this is not being changed by this
quantum revolution.  Computer scientists have a wide range of tasks
which they are interested in accomplishing successfully, safely,
and/or efficiently\cite{Brass}:
\begin{enumerate}
\item Given data $X$, compute $f(X)$ in the fewest number of steps. ({\em
computational complexity}\cite{compl})
\item Given two parties holding data $X$ and $Y$, compute $f(X,Y)$
with the least communication. ({\em communication
complexity}\cite{ccbook})
\item Given two parties holding data $X$ and $Y$, compute $f(X,Y)$ in
such a way that the two learn no more about each other's data than
they know from the function value itself. ({\em discreet function
evaluation} (\cite{Brass}, Chap. 5.8))
\item Transmit data $X$ reliably from one party to a second as quickly
as possible. ({\em channel capacity}\cite{Abr})
\item Protect data $X$ from duplication. ({\em counterfeit
protection})
\item Transmit data $X$ from one party to a second in such a way that
the data cannot be read by any third party. ({\em key
distribution/cryptography} (\cite{Brass}, Chap. 6))
\item Transmit data $X$ from one party to a second in such a way that
the receiver can be assured that the data was not corrupted during
passage through the channel. ({\em authentication}\cite{Brass})
\item Transmit data $X$ from one party to a second in such a way that
another party can later confirm that the second party did not alter
$X$, and can confirm that it was produced by the first party.  ({\em
digital signature}\cite{Brass})
\item Divide data $X$ among $n$ parties in such a way that no $n-1$ of
them can reconstruct data $X$, but all $n$ working together can. ({\em
secret sharing}\cite{sham})
\item Determine and execute optimal strategies in games. ({\em game
theory; economics}\cite{gamescl})
\end{enumerate}
In this information age, our society's well being increasingly depends
on being able to perform these and similar tasks well.  Quantum
mechanics is never mentioned in this list; nor should it be, since all
of these tasks involve the possession and transmission of data in
palpable, macroscopic form.  ``Quantum data'' is not useful for
members of our very macroscopic society; the inputs and outputs of
these tasks must be in classical form.  (One might question this
assumption in some radically altered definition of ``society''.)

But what we have increasingly realized is that the {\em tools}
employed to accomplish these tasks can well be quantum mechanical.  In
addition to ``classical'' processing primitives involved in completing
tasks (place a bit in memory, compute the AND of two bits, launch a
bit into a communication channel), we can employ a host of quantum
processing primitives: prepare a qubit (two-level quantum system) in
a particular pure state; launch a qubit (e.g., a photon or an electron,
see Sec. \ref{communication})
into a communications channel; transform the state of two qubits
according to the action of some two-body Hamiltonian.

The remarkable fact is that it is known how to achieve improvements in
many (but by no means all) of the tasks mentioned above by employing
these quantum primitives.  We will review here briefly the ``quantum
state of the art'' for our list of tasks:
\begin{enumerate}

\item {\em computational complexity:} Shor's famous work\cite{PS}
showed that some very important computations, for example prime
factorization, have only polynomial complexity if quantum primitives
are used, while this computation can (probably) not be done in
polynomial time if only classical primitives are used.  It is worth
reviewing the general way in which the classical specification of the
problem is converted into an application of quantum primitives: the
data $X$ (the number to be factored) is converted into a
time-dependent two-body Hamiltonian function which is applied to a set
of qubits prepared in a standard quantum state (e.g., all zeros).
Then the answer $f(X)$ (the set of prime factors) is obtained by the
results of a quantum measurement performed on each of the qubits.  It
may be necessary to repeat Shor's procedure several times to obtain a
factor.

It should be noted that there are some other computations for which it
has been proved that {\em no} improvement in computational complexity
is achieved by using quantum primitives\cite{polys}.  For instance,
the $n^{th}$ iterate of a function provided as a look-up table takes
$n$ references to the table even if quantum primitives can be
used\cite{Ozhi}.  Work continues to explore the cases in which quantum
speed-ups are and are not possible\cite{moreOzhi}.

\item {\em communication complexity:} In this work the advantages
gained by communicating using qubits rather than bits have been
explored\cite{CB}.  There are some strong positive results in this
area.  Quantum communication is provably more efficient for the
problem of two-party appointment scheduling: two persons have to
compare their appointment books to choose a day to have lunch out of
$N$ possible days.  For classical bit transmission $O(N)$ bits of
communication are required in general.  But it has been proved (it is
an application of the ``Grover'' algorithm\cite{Gro}) that no more
than $O(\sqrt{N}\log{N})$ quantum bits of transmission are needed to
complete this task with high reliability\cite{cleve98}.  There is a
related task in which the quantum speedup is even more dramatic, in
the area of ``sampling complexity'': two parties must both pick a
subset of cardinality $\sqrt{N}$ from a common set of size $N$ in such
a way that their subsets are disjoint.  Classically, $O(N)$ bits of
communication are required to assure disjointness, but just $O(\log
N)$ of quantum bit transmission suffices\cite{Ambai}.  Such dramatic
provable speedups are apparently also possible even in a case where
two parties share a string of random bits\cite{Raz}.

\item {\em discreet function evaluation}: This is an example of a
category of task for which there is believed to be {\em no} quantum
solution.  This is true, at least, for the principal technique which
computer scientists have used to analyze this task\cite{Brass}, which
involves reducing it to a procedure called {\em bit commitment}, in
which one party records a bit value of her choosing, locks the record
in a safe and sends it to a second party (the ``commit'' phase); then
at a later time of her choosing, she sends the key to the other party
(the ``opening'' phase).  Since safes can be x-rayed and locks picked,
this protocol is not secure.  It has been proved that bit commitment
is never secure in a quantum world\cite{bc}; using entanglement, the
sender can change the value of the bit between the commit phase and
the opening phase.

\item {\em channel capacity:} Here the results are tantalizing, but
not conclusive.  The problem is this: given a classical bit channel
and a quantum bit channel with the same levels of noise (the same
probability that the bit will pass through the channel unaffected,
roughly speaking), are fewer uses of the qubit channel needed to send
a given classical message reliably than of the bit channel?  No case
has been found in which this ``classical capacity of a quantum
channel'' exceeds the Shannon capacity for the classical channel,
although the work of Fuchs {\em et al.} give indications that it may
be possible\cite{Fuchs}.  Actually, there is one scenario in which the
quantum capacity is definitely greater: if the sender and receiver
have shared a prior supply of maximally entangled quantum states,
which themselves carry no classical information, the quantum capacity
can be boosted by the technique of superdense coding\cite{superd} by a
factor of two or sometimes more (at least up to a factor of three for
qubit transmission)\cite{priv}.

\item {\em counterfeit protection:} There are really no strong
classical techniques for protecting against counterfeiting.  The first
application of quantum primitives ever conceived, ``quantum money''
was devised in 1970 by S. Wiesner\cite{qmoney}.  It is a beautifully
straightforward application of the simple rules of quantum state
preparation and measurement.  The bank embeds qubits into its
banknotes; each qubit is in a pure quantum state, but the states are
drawn from a non-orthogonal ensemble (e.g., $|0\rangle$, $|1\rangle$,
$|0\rangle+|1\rangle$, and $|0\rangle-|1\rangle$, or in spin language,
in the eigenstates of either $\sigma_z$ or $\sigma_x$).  A record of
the state preparation is kept at the bank, and the bill is sent into
circulation.  When the note returns to the bank, the bank can use its
record to measure each qubit in a ``non-demolition''\cite{Caves}
fashion, that is, in the appropriate $\sigma_z$ or $\sigma_x$ basis so
that the state is undisturbed and the measurement outcome is
deterministic.  If all measurements agree with the stored record, the
bank can be assured that no attempt has been made by a counterfeiter
to read the state of the qubits to duplicate them.  This application
has not received much attention lately, but perhaps its day will come
with the further advance of quantum technology, when qubits can be
stored (or error-corrected) over very long times.

\item {\em key distribution:} The most well-known success of quantum
protocols is in ``quantum cryptography\cite{BB84}.'' The security of
quantum transmission of random data (the key) begins with the same
trick that is introduced in quantum money, sending one of a set of
non-orthogonal quantum states that an eavesdropper cannot reliably
distinguish, and that are in fact disturbed if the eavesdropper
attempts to learn any information about them.  The construction of a
secure key from this primitive involves a lot more work, but Mayers
has given a proof\cite{BB84sec} that the a protocol naturally obtained
from the one proposed by Bennett and Brassard in 1984\cite{BB84} is
unconditionally secure.  Another protocol in which state transmission
is augmented by local quantum computation is considerably easier to
prove secure\cite{QPAsec}.

\item {\em authentication:} Wegman and Carter\cite{Weg} introduced a
provably secure authentication technique that assumes that the sender
and receiver possess a secret key; therefore, a secure key exchange
using quantum primitives leads directly to a way of doing secure
authentication.  In today's world there is another way to perform
authentication: authentication is implied by digital signatures, which
are routinely used in present-day cryptography, but ---

\item {\em digital signatures:} The existence of quantum protocols has
negated the ability to do digital signatures.  First, no quantum
protocol can apparently be introduced which can take the place of
digital signatures used in public-key cryptography, in which a sender,
by appending to the end of a message an encrypted version of that
message, produces unalterable evidence that this message originated
from him: anyone can later decrypt the ``signature'' using the
sender's public key and compare it with the putative
message\cite{Brass}.  Second, the ``proof'' that this protocol is
secure relies on the security of public-key cryptography, which is
jeopardized by the ability to factor large numbers by quantum
computation.  Perhaps some entirely different quantum reasoning will
again permit the accomplishment of this information processing task.

\item {\em secret sharing:} Only a little work has been done on
this\cite{g8etal}, but it appears that there will be a variety of ways
of using multipartite states to split up a secret in such a way that
it can only be reconstructed by the cooperative quantum operations of
several parties.  Buzek {\em et al.} have shown ways in which this
problem can be approached using entangled states; it is perhaps more
surprising that it is possible to use unentangled quantum states to
perform this task.  This arises from the recent discovery that there
exist ensembles of multiparty orthogonal product states which can
nevertheless not be distinguished by any local operations of those
parties, even if they are allowed any amount of classical
communication.  Only a joint quantum measurement can distinguish them
reliably.  The detailed application of this discovery to a secure
secret sharing protocol has only just begun.

\item {\em games:} This is a rather ill-defined area at the moment,
but one with apparent promise.  Meyer and Eisert {\em et
al.}\cite{games} have shown that if the players in a game can perform
quantum mechanical manipulations in the game (e.g., moving a chess
piece into a superposition of positions by a unitary operation), they
can gain some advantages.  It seems that some changes will have to
take place in our society before some of these game results become
applicable --- can we have a quantum stock market?  A quantum economy?
\end{enumerate}

A final comment about this survey: while in some sense it covers
everything that goes on in the research on quantum improvements of
information processing tasks, in another way it misses a lot of what
workers in this field really think about.  Between the bottom level of
quantum or classical primitives like data transmission and qubit
measurement, and the top level of tasks to be accomplished, lies a
whole realm of macros and subprocedures which use the primitives and
provide tools for accomplishing the end tasks.  We are very familiar
with these in classical computing (fetch program instructions, invoke
a floating-point multiplier, launch a packet onto an ethernet), but
there is a whole host of quantum macros which have no classical analog
and which are crucial for facilitating the quantum implementations of
many tasks.  

An important example of these is quantum error correction and
fault-tolerant quantum computation\cite{Preskill}, which put together
the primitives of state preparation, measurement, and manipulation in
such a way that the effective unitary evolution of a quantum
computation is carried out reliably despite the intervention of noise
(``quantum decoherence'').  Another operation which one might consider
as a quantum macro is the sharing of a {\em quantum} secret, recently
discussed in \cite{CGL}.  Reliable qubit communication depends on
other noise-suppression quantum macros; the most effective approach to
this problem involves entanglement purification\cite{BDSW} (in which a
large supply of partially entangled mixed quantum states is
manipulated locally to produce a smaller supply of pure, maximally
entangled quantum states).  Another crucial subprocedure in this
noise-suppression macro is the celebrated ``quantum
teleportation\cite{bbcjpw}.''  Much of the recent ``Star Trek''
discussion of teleportation misses the point that it has a
well-defined, scientifically valid role as an an enabler of high-level
quantum processing for anything involving the transmission of quantum
information (e.g., distributed computing, key distribution).

So, quantum information processing isn't just factoring!  Quantum
factoring alone is interesting and important; seeing the whole
picture, though, indicates that we may be just at the beginning of
something {\em really} big.

\section{Quantum information processing and magnetic physics}

Specially-crafted magnetic materials and magnetoelectronic structures,
we believe, are good candidates for providing some of the important
primitive quantum tools for performing many of the tasks itemized
above, as we will detail shortly.  We will concentrate in this section
on
those applications which require the creation and manipulation of
``fixed'' qubits, which include the applications of quantum computing,
counterfeit protection, and secret sharing, and pieces of the others,
such as the encoding and decoding required in channel transmission.
In Sec. \ref{communication} we
will discuss a particular scenario based on mobile electrons
\cite{LBS}  whose spins provide the ``mobile'' qubits
needed in the other applications; some proposals are now being
considered in which the coupling by solid-state optical cavities to
photons \cite{Atac} could provide the tools for the remainder of our
tasks as well.

The magnetic structures that we envision are promising because the
qubit is naturally defined (in terms of a localized single spin).
This localized spin has the potential for being relatively well
isolated from its environment -- that is, for having low decoherence
rates -- and it can be manipulated by electrical, magnetic and/or
spectroscopic tools and can be measured using advanced magnetometric
or electronic techniques.

Of course, the magnetoelectronic structures that we propose are not
the only possible approach to the realization of quantum information
processing: efforts spanning many of the active areas of experimental
quantum physics have led to successful demonstrations of quantum logic
gates, and of operating systems for quantum cryptography, superdense
coding, and quantum teleportation.

We can only give a brief mention of all the different quantum logic
gate demonstrations that have been reported: In 1995, there was the
demonstration of the two-qubit controlled-NOT reported using ion trap
spectroscopy by the NIST group\cite{NIST}.  Since this demonstration,
progress towards realizing the idea of the linear-ion trap quantum
computer has been proceeding steadily; this group has recently
demonstrated the deterministic creation of entanglement between two
ions\cite{NIST2}.  In the area of cavity-quantum-electrodynamics, the
vacuum cavity version of the solid state microcavity scheme mentioned
above was first investigated in 1995 by the Cal Tech group\cite{CIT},
and many proposals have been made for how to use this device in a
quantum communication network.

The processing of photons in fiber-optic experiments has also received
a lot of attention.  Full-scale quantum cryptography demonstrations
have now been achieved in many different laboratories\cite{QC}.  In
addition, several other quantum information processing protocols have
been realized in such systems: superdense coding has been achieved in
systems where photon EPR pairs are created by parametric down
conversion, and incomplete Bell measurements are performed using
linear optical elements\cite{SD}.  More recently, teleportation of
photon polarization states has been achieved\cite{rometp,innsbrucktp}.
Now it has also become possible to teleport a ``continuous'' Hilbert
space, the quadrature field coordinates of a coherent state of
light\cite{cittp}.

Finally, it should be mentioned that there is another condensed matter
implementation of quantum gates that has received a lot of
experimenatal attention lately, one involving bulk NMR (nuclear
magnetic resonance).  Following on the original theoretical idea for
using NMR for quantum gates\cite{Div95}, the idea was put into
practical, realizable form in 1997\cite{NMR}.  Since then, there has
been a plethora of experimental investigations of 2 and 3 spin
systems, including demonstrations of the Deutsch-Jozsa and Grover
quantum algorithms\cite{Ike} and of simple quantum error correction
techniques\cite{Cory}.  There has even been a realization of
intramolecular quantum teleportation\cite{nmrtp}.

\subsection{Proposed Device Structure}

Rather than giving a general discussion of the criteria which a
magnetoelectronic device proposal must satisfy in order to be a good
candidate for a quantum computer (which we have done previously; see
Refs. \cite{the5,the6,the7}), we will simply proceed to describe the
specific model that we have introduced\cite{the6,the7,the8}.  From the
discussion here it should be clear what are the critical requirements
for this proposal to succeed.

Fig. 1 sketches the model that we have introduced in Ref. \cite{the6}.
It is a quantum-dot array\cite{jacak,kouwenhoven},
produced in this version of the
model by lateral confinement.  The figure indicates a two-dimensional
layer, for example a quantum well produced in a GaAs heterostructure,
above which an array of electrodes is placed.  As in the experiments
of various
groups\cite{ashoori,kotthaus,tarucha,waugh,oosterkamp,blick},
voltages on these electrodes can be used to deplete selectively
regions of the two-dimensional electron gas below them, leaving
isolated regions (the quantum dots shown dotted in the figure) in
which electrons can be confined.

The qubit in this scheme is provided by the {\em electron spin} of
each quantum dot.  In order that this qubit be well defined, the
electron number must be controlled and constant throughout the
operation of the device.  This is assured by exploiting the well
understood Coulomb blockade effect in these dots.  We imagine that a
transport (e.g., \cite{tarucha}) or capacitance\cite{ashoori}
measurement is performed on every dot in the array separately, and the
gate voltages (the gates are shown shaded in the figure) adjusted so
that the energy of the $N$ electron state is much lower than that of
the $N-1$ or $N+1$ electron state.  $N$ will remain fixed throughout
the quantum computation operations: this computer has no moving parts,
not even the electrons move (at least not much).  In order to use the
electron spin as a quantum number, it is very likely essential that
$N$ be an odd number (if $N$ is even, it would typically be the case
that the total spin of the dot would be zero, so that no
nearly-degenerate levels would be available to represent the qubit).
If $N$ is odd, the spin is at least $1/2$.  In fact, $s=1/2$ exactly
is the ideal situation for representing a qubit.  $s=1/2$ is assured
if $N=1$, that is, if there is only one excess electron confined to
the dot.  For this reason partly, but mainly because of other
considerations about the many-body physics of the dot, such as that
discussed in Sec. III, we will consider only the $N=1$ case.  The
$N>1$ case may be usable for quantum computation, but it will require
more analysis than we have performed up until now.

$N=1$ is not easy to achieve experimentally.  $N$ in the range of a
few tens has become relatively routine in the experiments cited, but
in the very small-$N$ regime it becomes difficult for electrons to
tunnel in and out of the dot, and the quantum-dot potential can become
disorder dominated.  These are not severe difficulties in principle,
but we acknowledge that it is a demanding requirement from the
perspective of present-day experiments, and we are committed to
studying the effect of using larger electron numbers on our proposed
device operation.

\subsection{Decoherence}

Among the most crucial requirements for the implementation of quantum
logic devices is a high degree of quantum coherence.  Coherence is
lost when a qubit interacts with other quantum degrees of freedom in
its environment and becomes entangled with them.  Predicting the
coherence time of the electron spin states of the device described
above is very difficult, as the possible couplings to all the other
quantum degrees of freedom of the system must be considered.  We are
encouraged, however, by the general fact, observed in many
experimental situations in condensed matter physics, that spin degrees
of freedom have longer coherence times than charge degrees of freedom
(ones for which the different electron states are associated with
different orbital wavefunctions), simply due to the weaker couplings
of spin states than orbital states to the environment.

This observation does not lead to any simple result about what the
available decoherence times in our structure will be.  Experiments of
spin coherence times have been performed on somewhat related
structures\cite{Kik,GAPA}, with the result that a very wide range of
decoherence times can be seen for the spins of electrons in
semiconductor heterostructures and bulk doped semiconductors.  In
structures which are intentionally doped with magnetic ions (Mn), the
coherence times are seen to be very small, on the order of
picoseconds.  But times ranging over six orders of magnitude,
approaching microseconds in some structures, have now been seen
depending on the details of the semiconductor structure.  As we will
discuss in the next section, microsecond decoherence times would be
acceptable for beginning experiments on quantum gate operations, while
times of milliseconds would be adequate for even large-scale quantum
computing applications (because of the abilities offered by quantum
error correction\cite{Preskill}).

We have considered in general the likely mechanisms of decoherence in
structures such as Fig. 1, which should be useful in guiding designs
of experiments which seek to lengthen the decoherence times.  Our
estimates\cite{the8} indicate that decoherence due to spin-orbit
coupling should be negligible for conduction band electrons in GaAs
(although not for holes); still, more detailed work needs to be done
to quantify this effect.  A potentially important mechanism for
decoherence in these structures is the coupling to other spin states
in the environment.  As demonstrated in the Mn-doping
experiments\cite{Kik}, this effect will be greatly influenced by the
materials preparation of the devices, and can be a very strong pathway
to decoherence.  Thus, in our work\cite{the6} we have studied in
detail models in which the qubits are coupled to a bath of other
spins.  The significance of the effect is entirely determined by the
strengths of the coupling constants between the system and bath.  The
decohering effect of this bath can be enhanced during quantum gate
operations\cite{the7}, that is, when spins in neighboring quantum dots
are coupled (see next section).

In addition to other electronic spins, there are unquestionably
nuclear spins in the environment as well whose decohering effect must
be considered.  In GaAs in particular, 100\% of the nuclei possess
non-zero spin.  We have studied the effect of these spins
recently\cite{the8}, and our calculations indicate that these spins can
be a serious source of decoherence if the applied magnetic fields are
low and the nuclear spins are in their thermal equilibrium state.
However, it is relatively easy to modify these conditions, either by
dynamically spin-polarizing the nuclear spins e.g. by known optical
techniques, and/or by arranging that the operation of the devices is
performed with a non-zero applied magnetic field.  Actually, the presence
of significantly spin-polarized nuclei may actually be very useful for
performing gate operations on these qubits (see next
section)\cite{the8}.

Another ``trivial'' but practically important source of spin
decoherence arises from uncertainties in the applied Hamiltonians to
be discussed in the next section.  For example, in schemes in which
the gate action involves the application of a uniform magnetic field,
inhomogeneities in this field will result in inaccuracies in the gate
operation.  This decoherence effect is analogous to the broadening
effect on absorption lines which is well known in traditional spin
spectroscopies, where various ``refocusing'' and ``spin-echo''
techniques have been devised to ameliorate them.  Such techniques may
have to be developed and adapted to assure reliable quantum gate
operation, but this problem has not been addressed systematically in
any detail.

One might think that if fluctuating magnetic fields are a severe
problem for quantum-dot quantum bits, then perhaps there would be some
value in reconsidering the use of electron orbital states, which,
after all, would be insensitive to such magnetic field effects.  We
are pessimistic on this account, not only because the decoherence
times for orbital states are short for myriad other reasons, but also
because there is reason to believe that some of the important
decoherence mechanisms due to Fermi-sea effects will be non-Markovian.
Markovian, or memoryless, decoherence is actually greatly desired over
non-Markovian decoherence in quantum computation, as all the powerful
techniques introduced in quantum error correction assume a memoryless
error scenario\cite{Preskill}.  No one has demonstrated that a qubit
system with even very weak non-Markovian decoherence would be useful
for quantum information processing.

\subsection{Quantum gates}

Another crucial requirement for quantum computing, and for many of the
other quantum approaches to information processing tasks outlined in
the first section, is that it must be possible to apply time-dependent
one- and two-body Hamiltonians to the qubits according to the
specifications of some program\cite{crock}.

The structure of Fig. 1 can have many mechanisms for applying such
``quantum gates'' to the spin qubits.  First, the structure has a set
of gates which can control the position of the electron's wavefunction
within the two-dimensional electron gas, simply by varying the
confining voltages on these gates.  If two of these electrons in
neighboring dots are pushed close together, the overlap of the orbital
wavefunctions will, via the Pauli principle, produce an effective
two-spin interaction between the two spin qubits.  The Hamiltonian
produced is that of an exchange interaction which is isotropic in spin
space
\begin{equation}
H(t)=J(t){\bf{S}}_1\cdot{\bf{S}}_2.
\label{heisenberg}
\end{equation}
Here the time dependence $J(t)$ is regulated by the time variation of
the tunneling matrix element $\Gamma$ of an electron from one dot to
the other.  According to perturbation theory, $J(t)$ is
\begin{equation}
J(t)\propto{\Gamma(t)^2\over U}.\label{exch}
\end{equation}
Here $U$ is the Coulomb blockade energy, the charging energy required
to add a second electron to one of the dots.

In Ref. \cite{the8} we give a more refined and detailed analysis of
this switchable spin interaction, in particular we show that the long
range part of the Coulomb interaction (if it is not screened) will
produce an additional term in (\ref{exch}) of opposite sign that leads
to a sign reversal of {\em J} for sufficiently large external magnetic
fields as a result of competition between long-range Coulomb repulsion
and magnetic wave function compression. By working at this magnetic
field (where {\em J} vanishes) the exchange interaction can be pulsed
on, even without changing the tunneling barrier between the dots,
either by an application of a local magnetic field, or by exploiting a
Stark electric field (which will also make the exchange interaction
nonzero).  See \cite{the8} for further information.  We finally note
that the exchange energy {\em J} can be understood as the level
splitting induced by the formation of a molecular state between the
two quantum dots\cite{the8}. The observation of such a molecular state
in a double dot system containing several electrons has indeed been
reported recently \cite{blick,Oosterkampmolecule}.

The exchange interaction of the form Eq. (\ref{heisenberg}) is sufficient
for the most general quantum computation, if it is supplemented by a
suite of one-body time-dependent interactions (one-bit gates).  This
is discussed in \cite{the6,the7,the8}, where it is shown that
Eq. (\ref{heisenberg}) will produce a quantum gate known as a
``square-root of swap'' (in which the exchange interaction is turned on
for half the time required for it to produce a complete interchange
(``swap") of the quantum states of the two qubits).  We show \cite{the6}
that two square roots of swap, in conjuction with a set of one-qubit
gates, will produce a quantum XOR (also known as a controlled-NOT) gate,
which is known to be employable for any arbitrary quantum
computation\cite{Barenco}.

The speed at which these switchings are done will be an important
parameter; the rule is, the faster the better, consistent with doing
the prescribed manipulations with rather high accuracy (error
correction theory says that the relative accuracy to be striven for is
on the order of $10^{-4}$).  The fundamental physics says that the
switching on and off of the tunneling could be done much faster than a
nanosecond\cite{the8}--only at much, much shorter time scales will such
fundamental limitations as adiabaticity enter the picture.  It is
necessary that the switching time be smaller than the decoherence
time; again, error correction theory says that ultimately, it is
desirable that the switching time be smaller than the decoherence time
by about $10^{-4}$.  We think that $10^{-1}$ will be quite
satisfactory for the initial round of measurements.  We think that
initially, the experimentalist should simply be guided by what is
doable.  Since high-frequency signals are difficult to transmit into
quantum dot structures in the Coulomb blockade regime at 4K or so, we
might suggest that one should shoot for switching times in the
neighborhood of $10^{-7}$ sec.  A simple calculation indicates that
only modest control-voltage excursions are needed to do
square-root-of-swap in this time.

We note that the switching of the gates via an external control field
$v(t)$ should be performed adiabatically \cite{the8},
i.e. $|\dot{v}/v|\ll \delta \epsilon/\hbar$, where $\delta \epsilon$
is a characteristic energy scale of the problem. In the present case
$\delta \epsilon$ should be taken to be on the order of the orbital
energy-level separation.  This adiabaticity requirement excludes e.g.
switching pulses of rectangular shape, in which case many excitations
into higher levels will occur. An adiabatic pulse shape of amplitude
$v_0$ is e.g.  given by $v(t)=v_0 {\rm sech}(t/\Delta t)$, where
$\Delta t=\tau_s/\alpha$ gives the width of the curve and $\alpha$ is
chosen such that $v(t=\tau_s)/v_0$ becomes vanishingly small.  In this
case we have $|\dot{v}/v|={1\over \Delta t}\,|{\rm tanh}(t/\Delta
t)|\,\leq {1/\Delta t}=\alpha/\tau_s$, and thus for adiabaticity we
need to choose $\tau_s$ such that $\alpha/\tau_s\ll \delta
\epsilon/\hbar$.  Note that the Fourier transform, $v(\omega)=\Delta t
v_0 \pi {\rm sech}(\pi\omega\Delta t)$, has the same shape as $v(t)$
but with a width $2/\pi\Delta t$, and we see that $v(\omega)$ decays
exponentially in frequency $\omega$, whereas it decays only as
$1/\omega$ for a rectangular pulse.  [We could, of course, also use a
Gaussian pulse shape, however, in this case we would get
$|\dot{v}/v|\propto t$ and some cutting of the long-time tails is
required in order to satisfy adiabaticity for all times.]  It is worth
emphasizing, however, that for our quantum gate action the pulse shape
is not relevant, the only parameter which counts is the integrated
pulse shape, $\int_0^{\tau_s} dt P(v(t))$, where $P$ stands for the
exchange $J$ or the magnetic field $B$ which is switched. This stands
in contrast to spectroscopic mechanisms based on resonance conditions
where more details of the shape of the pulse are relevant.  Also, even
if adiabaticity is not well satisfied in our switching, not much will
happen as long as spin-orbit coupling remains small since typically
only charge degrees of freedom will be excited in a non-adiabatic
process and not the spins representing the qubits.

The use of an inhomogeneous magnetic field (or an inhomogeneous
g-factor) for gating mentioned above for two-bit gates is obligatory,
in some form, for the accomplishment of the desired one-qubit gates.
That is, every one-body Hamiltonian needed for quantum computing can
be written in a standard Zeeman form
\begin{equation}
g\mu_B{\bf B}(t)\cdot{\bf {S}}.
\end{equation}
It is necessary that the field ${\bf B}(t)$ (or the effective field)
be applicable separately to each qubit (or at least that the effect
on neighboring qubits be smaller and known), and that it can be applied
along at least two different axes.

There are many ways that we can conceive of applying these local
magnetic fields or local Zeeman interactions.  If the switching time
scale is to be the same as above ($10^{-7}$ sec.), then field
strengths of only a few Gauss are necessary, and this could be
accomplished by a mechanism as simple as winding a small wire coil or
by placing magnetic dots above/below each quantum dot, or by placing
the dots between a grid of current-carrying wires as in RAM devices
\cite{Prinz}.  Other methods of obtaining very localized fields, such
as moving magnetic bubbles in a garnet film, using a magnetic-disk
writing head, or a magnetic force microscope tip, can be considered.

Although strict localization of the applied field is not necessary, it
does make life considerably easier, and there are several ideas which
would make this field effectively much more localized.  If the nuclear
spins of the dot and the material surrounding the dot can be polarized
as discussed above, then the electron spin (but not the orbital
motion) experiences an effective internal magnetic field, the
``Overhauser field'', which can be on the order of several Tesla in
GaAs\cite{the8}.  If the Overhauser field is different in the dot and
in the confining layers above and below it, then the field as seen by
the confined electron can be varied by purely electric gating, that
is, by pushing the electron more or less into the insulating barriers.
In our original work\cite{the6} we introduced another variant of this
idea, in which the confining materials possess a {\em real}
magnetization due to a ferromagnetic moment.  Such ferromagnetic
insulating materials are not so common, but are not unheard of either
(the garnets, the ferrites, and the Eu-chalcogenides are some
examples); unfortunately, there is little experience in matching these
materials epitaxially to the common dot materials such as GaAs, but
first promising progress in this direction has been made recently, see
\cite{Kleiber}.  We also would like to emphasize here that our set-up
permits the performance of swaps of qubit states in such a way that we can
easily move a spin state (not the electron spin itself) of a given quantum
dot via a chain of adjacent quantum dots to a desired location in the
network where we have localized magnetic fields available, act with the
field on the qubit and then swap the qubit back to its original location.
This is possible since the swapping operation does not involve
single-qubit rotations and since we can swap two states even without
knowing their particular state.  But either the Overhauser field idea or
the magnetic insulator idea can be extended to solve the very important
problem of quantum measurement, to be discussed momentarily.

A brief word about error correction, which we have alluded to many
times already: error correction provides a way of using redundancy and
repeated quantum measurement during the course of computation, which
detects and diagnoses the occurrence of decoherence, and undoes its
effects.  It uses exactly the same gates which we have just
introduced, along with qubit measurements to be described shortly.
The conventional analysis of quantum error correction\cite{Preskill}
assumes that two-qubit gates can be performed between any two qubits.
In our computational model, gate operations can only be performed
between neighboring qubits.  This is not a serious modification, the
basic procedures of quantum error correction still work in this
case\cite{Gott}.  The more crucial requirement for error correction to
work is that two-qubit and one-qubit gates can be performed on many
different qubits simultaneously as it is possible in our proposal.
There are other popular quantum
register designs, for example the well-known linear ion trap model of
Cirac and Zoller\cite{CZ}, for which error correction is not possible
because gate operations cannot be done in parallel.

Finally, the concept of error correction promises to be important by
itself. Indeed, in many areas of mesoscopic physics it would be highly
desirable to maintain phase coherence {\it indefinitely}, a goal
which we believe could be achieved with error correction schemes.

\subsection{Quantum measurements}

The final requirement which must be addressed for performing quantum
information processing with the quantum-dot structure is the need to
read out data reliably, which translates into the necessity of doing
spin measurements at the single-spin level.  It must be possible to
address each individual spin in the structure (or at least some subset
of the spins) and perform an ``up/down'' measurement on them.
Solid-state magnetometry at the single-Bohr-magneton level has of
course proved to be very difficult, as other contributions to this
volume will discuss.  We forsee, though, that using some of the
capabilities of quantum computing, the very difficult single-spin
measurement can be turned into a more manageable electrical (i.e.,
charge) measurement along the lines first proposed by us in
Ref. \cite{the6}.

We have recently reviewed in detail the possibilities in this area, we
will just give an outline here, the interested reader is referred to
\cite{spinf}.  The basic idea of turning the spin measurement into a
charge measurement \cite{the6} (see also \cite{Kane}) is this: we use the
kind of magnetic (either ferromagnetic or nuclear-spin-polarized) barriers
mentioned above as tunnel barriers, say in the form of a thin barrier
separating two quantum dots or a quantum dot and a single-electron
transistor.  The tunneling barrier can be made strongly spin dependent
(this is the well-known ``spin-filter'' effect); thus, at the time of
measurement, the tunneling of a spin-up electron can be made very
probable, while the tunneling of a spin-down electron remains very
improbable.  Thus, the job of measuring spin is converted into the job
of measuring whether an electron has tunneled or not.  But this is a
feasible (and indeed, almost routine) electrometry measurement--many
labs have demonstrated the feasibility of single-electron-charge
magnetometry, either with single-electron transistors, quantum point
contacts, and other mesoscopic electronic structures.

Another promising idea for single-spin measurement involves near-field
optical probing of the spin state.  We have not analyzed this approach
in any detail, but it deserves future experimental and theoretical
attention.

\subsection{Test experiments}

It is clear that the above concept, which we have developed over the
last three years, has proved far too demanding to be undertaken all at
once.  It requires a combination of developments, in materials and
device fabrication, in precision, high frequency electrical control,
in hitherto unexplored, complex, nanoscale architectures, which are
far beyond the scope of one generation of experimental investigation.

Therefore, it is very important to pull apart our quantum-dot quantum
computer into small pieces, setting feasible shorter-term goals for
the demonstration of particular capabilities.  We only intend to give
a brief idea here of the kind of near-term work which might be done;
indeed, it seems that the possible ways of dividing our proposal into
smaller, manageable chunks are almost infinite, and finding the most
promising ones can only result from a detailed dialog between the
theorist and experimentalist.  But here is a selection of ideas which
we now now might be promising for the next few years:

There is a clear need to demonstrate the controlled fabrication of
spin quantum dots.  As mentioned above, a desirable goal would be to
routinely obtain dots with just {\em one} excess electron.  More theory
must be done to see whether using dots with an odd number of excess
electrons would be acceptable.  One-electron dots have been achieved
\cite{ashoori}, but not in geometries in which dots could potentially
be coupled.  Loading by transport in the Coulomb blockade regime would
be the obvious way, but doping or optical techniques should also be
considered.

If an array of such dots can be obtained, then characterization of the
qubit energy levels, g-factors, and especially decoherence times would
be the next thing to study.  In fact, an initial version of this type
of experiment has now been reported\cite{GAPA}, which demonstrates
that time-resolved optical probes of these systems are extremely
promising for these kinds of initial characterizations.  Further
application of pulsed-spectroscopy techniques should yield further
information about the controllability of such qubits (at least at the
one-qubit gate level).

Another distinct line of investigation would involve demonstration of
two-qubit gate capabilities.  We have suggested\cite{the6,the7,the8}
that gated double-dot structures that have been fabricated and studied
in GaAs 2DEGs\cite{tarucha,waugh} could be the starting point of such
studies; it will also be desirable to see if other types of dots, say
in pillar structures or ones created by chemical nucleation, can be
integrated into devices in which their coupling is subject to
electrical or magnetic control.  We envision experiments in which arrays
of these dots can be subjected to identical preparations and probings.  It
may be that an experiment as straightforward as the measurement of the
a.c. magnetic susceptibility of such a dot array as a function of a
control voltage\cite{the6,the7} will be sufficient to demonstrate the
basic physics of quantum-mechanical exchange coupling between
neighboring spins.

The magnetoelectronic techniques that we have suggested for other gate
operations and for single-spin quantum measurement involved additional
and quite different experimental challenges.  The basic materials
issues of the integration of semiconducting and magnetic materials are
not yet well enough developed to even propose a likely system to study
at this time, although it is promising to note that there is now
active research focussed on just this area, finding good matches
between magnets and semiconductors which will show clean, reproducible
interface properties.  If, for example, it proves possible to grow EuS
or EuO on GaAs, then an experiment can immediately be considered in
which the basic spin filtering phenomenon of carriers in the
semiconductor conduction band is looked for.  This experiment would be
very informative even in a traditional bulk tunneling geometry; there
would be no need to even consider integrating these with quantum dot
structures at first.  Tunneling through Overhauser-polarized barrier
materials may be less demanding from the materials science point of
view, but will require integration of optical (for nuclear spin
polarization) and electrical expertise.  A later generation of experiment
could consider integrating the spin-filter into a simple point-contact
(say of the Ralls type) so that a combined spin-filter/Coulomb blockade
effect could be demonstrated.  This already takes us quite far into
speculative territory.

We would like finally to briefly comment about questions that we have
been asked about whether the many experiments on the {\em charge}
degree of freedom in quantum dots could be directed towards the
achievement of orbital-level qubits and quantum gates.  While there
may be a worthwhile approach in this direction, we are pessimistic
about its ultimate chance of success compared with the spin approach,
even though spin effects are at this time much less well developed in
quantum-dot research.  We say this based on the fact that orbital (i.e.
charge)
degrees of freedom of a dot will be much harder to make coherent than
the spin of a dot, just based on the typically stronger coupling of
charge (compared to magnetic moment) to the environment.  A typical
Fermi-sea charge environment also has a different, and possibly even
worse, problem as already pointed out before: Fermionic baths are very
non-Markovian, having power-law decays of correlations. Almost all the
well-developed theory of quantum error correction applies only to
Markovian baths\cite{Preskill}, and it is very unclear whether any useful
quantum computation can be done in the presence of a non-Markovian
environment (however, see \cite{Viola}).  These considerations have been
enough to justify, in our minds, a continued focus on the eventual
possibilities of spin quantum dots only.

\section{Quantum Communication with electrons}
\label{communication}

In this section we would like to address the following question: is it
possible to use mobile electrons, prepared in a definite (entangled)
spin state, for the purpose of quantum communication? Such a question,
for instance, is of central importance in a solid state quantum
computer where one wishes to exchange quantum information between
distant parts of a quantum network.  The question is of course also of
broader interest: if we could use electrons for creating entangled
states, in particular so-called EPR pairs, and if we could move them
around separately while preserving their spin entanglement, then we
would be able to implement, for instance, tests of Bell's inequality;
thereby, we could obtain tests of non-locality---one of the most
striking concepts of quantum mechanics---for the first time with
electrons.
So
far, all such tests have been done on photons \cite{Aspect},
most recently by Gisin's group\cite{Gisin} who demonstrated
in a remarkable experiment
that photons propagating in optical
fibers remain in an entangled state
over more than 10 km's.
It is quite amusing to note here that the
Gedanken experiment which has been formulated by Einstein, Podolsky,
and Rosen\cite{Einstein}, and which underlies the Bell inequalities,
makes use of point particles and not of massless particles such as photons.
Thus, there can be no doubt that it would be highly desirable to
extend tests of non-locality also to quantities which have a rest mass
such as electrons in particular.

Now, as we have discussed before,
one basic ingredient for quantum communication are entangled pairs
of qubits which are shared by two parties.
There are three separate requirements involved here which must be satisfied.
First of all we need {\it mobile qubits} which can be transported
from position A to position B. Second, we need
a source of entanglement for such qubits
which can be operated in a controllable way, and third,
it must be possible to transport each of the qubits separately in a
phase-coherent manner such that the
entanglement between the two qubits of interest is not destroyed in the
process of transporting them to their desired locations.

Now, our choice of representing the qubit in terms of the spin of a
mobile electron satisfies the first requirement trivially (note that
qubits defined as pseudospins are typically not mobile).  The second
requirement, to have a source of entanglement, can be satisfied by
using the quantum gate mechanism based on coupled quantum dots
\cite{the6,the7,the8} as we have described it in the preceeding
sections.

To assess the third requirement, transport of entangled qubits, we
need to be more specific of how we actually envisage such transport.
One realistic scenario is to attach leads to the quantum dots into
which the electrons can be injected (e.g. by lowering the gate
barriers between dot and lead).  From an experimental point of view it
is best to make leads and dots out of the same material.  For
instance, if the dots are formed in a two-dimensional electron gas
(2DEG) such as GaAs heterostructures it is not difficult to connect
them to leads formed also in the 2DEG by electrostatic confinement or
some etching techniques \cite{blick,tarucha}.  In a first step we
inject an electron into quantum dot 1 and another one into quantum dot
2. In a second step, we perform a quantum gate operation to produce an
entangled state out of the two electrons, say a singlet state,
$|\psi_{kk'}\rangle={1 \over \sqrt{2}}(|k,k'\rangle + |k',k\rangle)
\chi_s$, where  $\chi_s=(|\uparrow\rangle_1|\downarrow\rangle_2
-|\downarrow\rangle_1|\uparrow\rangle_2)/\sqrt{2}$ is the two spinor
describing a spin-singlet state. The orbital part of the state,
characterized by the quantum numbers $k,k'$, is symmetric whereas the
spin singlet is antisymmetric. As a measure of correlations we
consider transition amplitudes between an initial and a final
state. We begin with the simplest case given by the wave function
overlap of $|\psi_{kk'}\rangle$ with $|\psi_{qq'}\rangle$,
\begin{equation}
\langle \psi_{qq'}|\psi_{kk'}\rangle=
\delta_{qk}\delta_{q'k'}+\delta_{qk'}\delta_{q'k} \, .
\label{overlap}
\end{equation}
Thus, if e.g. $q=k$, and $q'=k'$, the overlap assumes its maximum
value one, simply reflecting maximum correlation between the two
states.  If we prepare the two electrons in a triplet state instead of
a singlet we will find a minus sign instead of the plus sign in
Eq. (\ref{overlap}).  This means that this sign simply reflects the
symmetry of the orbital part of the wave function, and thus the
overlap (\ref{overlap}) distinguishes only triplet from singlet states
but not necessarily entangled from unentangled states. Indeed, the
triplet states with $m_z=\pm 1$ are not entangled, whereas the triplet
state with $m_z=0$ as well as the singlet state are entangled.  Since
the (anti-)symmetry of the orbital part of the wave function leads to
(anti-)bunching behavior in the noise spectrum \cite{Brown}, we can in
principle distinguish singlet from triplet states.  The triplet states
themselves can be further distinguished by measuring the {\em
z}-component of the total spin, $S_z$, which could be achieved e.g. by
making use of spin filters in the leads and/or leads that are
connected up to other quantum dots into which the electrons can tunnel
and then be detected via SET measurements\cite{LBS}. In this way it is
possible (in principle) to distinguish all four spin states, in
particular also to distinguish between entangled and unentangled
states (provided we deal with these four particular states
only--otherwise the expectation value of $S_z$ does not distinguish
between entangled and unentangled states in general).

Next we generalize this concept of
the overlap to a dynamical situation as well as to the leads
which contain many interacting electrons besides the two
entangled electrons of interest. Again, we  use a similar overlap
as a measure of how much weight remains in the final state
$|\psi_{qq'}, \psi_0, t\rangle$
when we start from some given initial state
$|\psi_{kk'},\psi_0 \rangle$, where $\psi_0$ denotes the fermionic
ground state of the electrons in the leads, which is simply given by a filled
Fermi sea.
For further discussion it is now convenient
to make use of the standard second quantization formalism
in terms of fermionic
creation ($a^\dagger_{k\sigma}$) and annihilation
($a_{k\sigma}$) operators, where $\sigma=\pm 1$ denotes spin $\uparrow$ ($\downarrow$)
in the $S_z$-basis.
The (normalized) initial  state, choosing a singlet,
can then be written as
\begin{equation}
|\psi_{kk'},\psi_0 \rangle={1\over \sqrt{2}}
(a_{k\uparrow}^\dagger a_{k'\downarrow}^\dagger-
a_{k\downarrow}^\dagger\, a_{k'\uparrow}^\dagger\,)\, |\psi_0\rangle \,\, ,
\end{equation}
and similarly for the final state, again chosen to be a singlet state.
The overlap (\ref{overlap}) now becomes a singlet-singlet
correlation function which we denote by $G^s(q',q,t;k,k')$, $t\geq 0$,
and which
is explicitly given by
\begin{eqnarray}
G^s(q',q,t;k,k')&=&{1\over 2} \sum_{\sigma=\pm 1}[ G(q',-\sigma;q,\sigma; t;
k,\sigma;k',-\sigma) \nonumber \\
&-& G(q',-\sigma;q,\sigma; t;
k,-\sigma;k',\sigma)]\,\, ,
\label{generaloverlap}
\end{eqnarray}
where
\begin{equation}
G(q',-\sigma;q,\sigma; t;
k,\sigma;k',-\sigma) =- \langle\, T\,
a_{q' \sigma}(t)\, a_{q -\sigma}(t)\,
a_{k -\sigma}^\dagger\, a_{k' \sigma}^\dagger\, \rangle
\,\,
\label{2Green}
\end{equation}
is a standard 2-particle Green's function, and $k=({\bf k},k_l)$,
where $k_l=\pm 1$ refers to lead 1 (2). Here, T is the time-ordering
operator and $\langle ...\rangle$ the zero-temperature or ground state
expectation value.  We assume a time- and spin-independent
Hamiltonian, $H=H_0+\sum_{i<j} V_{ij}$, where $H_0$ describes the free
motion of the $N$ electrons, and $V_{ij}$ is the bare Coulomb
interaction between electrons $i$ and $j$ (extensions to more
complicated situations including spin interactions will be considered
elsewhere).  This four-point correlation function is of the type
$G(12; 1'2')$ and it provides a measure of how much overlap (or
transition amplitude) is left after time {\em t} between an initial
and final singlet state of two electrons which have been injected into
a Fermi sea (leads) of $N-2$ interacting electrons, and which
propagate during time {\em t} in the leads before they are taken out
again.  We emphasize that after injection the two electrons of
interest are, of course, no longer distinguishable from the electrons
of the leads, and consequently the two electrons taken out of the
leads will, in general, not be the same as the ones injected.

It is now a non-trivial many-body problem to find an explicit value
for $G(12; 1'2')$. On the other hand, we can expect some
simplification: without spin-dependent forces we know that the total
spin must be conserved even if the two electrons strongly interact
with the rest (and among themselves) via Coulomb interaction.  It is
thus not unreasonable to expect that we still find some spin
correlations, in particular entanglement, between initial and final
states. But how much is it? And why and how do we loose some of the
correlations, etc.? These questions are of fundamental interest, and
we can find answers to them by evaluating $G(12; 1'2')$ explicitly
with the help of standard many-body techniques\cite{Fetter,LBS}.
Omitting most of the details\cite{LBS} here we briefly state the main
results. First we note that the four-point Green's function
considerably simplifies for the realistic situation where there is no
Coulomb interaction between the electrons in lead 1 and the electrons
in lead 2. As a result the 2-particle vertex part vanishes and we get
$G(12; 1'2')=G(11')G(22')-G(12')G(21')$, i.e. the Hartree-Fock
approximation is exact and the problem is reduced to the evaluation of
single-particle Green's functions $G_1({\bf k},t)$, $G_2({\bf k'},t)$
pertaining to lead 1 and 2, resp.  (these leads are still interacting
many-body systems though).  In particular, we now find
\begin{eqnarray}
G^{s/t}(q',q,t;k,k')= &-&
\{ G_1({\bf q},t)\, G_2({\bf q'},t)\, \delta_{{ qk}}\delta_{{ q'k'}}
\nonumber \\
&\pm& G_1({\bf q'},t)\, G_2({\bf q},t)\,
\delta_{{ qk'}}
\delta_{{ q'k}}\} \, ,
\end{eqnarray}
where the upper (lower) sign refers to the spin singlet (triplet),
and where we have chosen $k_l=1$.
For the special case $t=0$, $N=2$, and no interactions, we have $G_j=-i$,
and thus
$G^s$ reduces to the rhs of Eq. (\ref{overlap}). For the general case,
we evaluate
the (time-ordered) single-particle Green's functions $G_j$ close to
the Fermi surface
and get the standard result\cite{Fetter}
\begin{equation}
G_j({\bf q},t) \approx -i z_q
\Theta(\epsilon_q -\epsilon_F) e^{-i \epsilon_q t- \Gamma_qt}\,\, ,
\label{quasiparticlepole}
\end{equation}
where $\epsilon_q=q^2/2m$ is the quasiparticle energy (of our additional electron),
$\epsilon_F$ is the Fermi
energy, and $1/\Gamma_q$ is the
quasiparticle lifetime. In a 2DEG, $\Gamma_q \propto (\epsilon_q -\epsilon_F)^2
\log(\epsilon_q -\epsilon_F)$ \cite{Quinn} within the random phase
approximation (RPA), which accounts for screening and which is
obtained by summing all polarization diagrams\cite{Fetter}.  Thus, the
lifetime becomes infinite when the energy of the added electron
approaches $\epsilon_F$.  Eq. (\ref{quasiparticlepole}) is valid for
$0\leq t\lesssim 1/\Gamma_q$, in which case the incoherent part of the
Green's function is negligible.
 Now, we come to the most important quantity in the present context, the
renormalization factor or quasiparticle weight, $z_F=z_{q_F}$,
evaluated at the Fermi surface; it is defined by
\begin{equation}
z_F={1 \over {1-  {\partial \over
{ \partial \omega}}Re \Sigma( q_F,\omega=0)}}\, ,
\end{equation}
where $\Sigma( q,\omega)$ is the irreducible self-energy occurring in the
Dyson equation.  The quasiparticle weight, $0\leq z_q\leq 1$, describes
the weight of the bare electron in the quasiparticle state ${\bf q}$,
i.e. when we add an electron with energy $\epsilon_q \geq \epsilon_F$
to the system, some weight (given by $1-z_q$) of the original state
${\bf q}$ will be distributed among all the electrons due to the
Coulomb interaction. This rearrangement of the Fermi system due to
interactions happens very quickly, at a speed given approximately by
the plasmon velocity, which exceeds the Fermi velocity (typically
$10^5$ m/s in GaAs).  Restricting ourselves now to momenta close to the
Fermi surface and to identical leads (i.e. $G_1=G_2$) we then have
\begin{equation}
|G^{s/t}(q',q,t;k,k')|
=\, z_F^2 \, |\, \delta_{{ qk}}\delta_{{ q'k'}} \pm
\delta_{{ qk'}}\delta_{{ q'k}}|\,
\end{equation}
{\it for all times} satisfying $0< t\lesssim 1/\Gamma_q$. Thus we see
that it is the
quasiparticle weight squared, $z_F^2$, which is the measure
of our spin correlation function $G^s$ we were looking for.
It is thus interesting to evaluate $z_F$ explicitly. This is
indeed possible, again within
RPA, and we find after some calculation \cite{LBS}
\begin{equation}
z_F=1-r_s ({1\over 2} +{1\over \pi})\,\, ,
\label{qpweight}
\end{equation}
in leading order of the interaction parameter $r_s=1/q_F a_B$, where
$a_B=\epsilon_0\hbar^2/me^2$ is the Bohr radius.  In particular, in a
GaAs 2DEG we have $a_B=10.3$ nm, and $r_s=0.614$, and thus we
obtain from (\ref{qpweight}) the value $z_F=0.665$. We note that a more
accurate numerical evaluation of the exact RPA self-energy yields
$z_F=0.691155$\cite{LBS}, again for GaAs.
[For 3D
metallic leads with say $r_s=2$ (e.g. $r_s^{Cu}=2.67$) the loss of
correlation is somewhat less strong, since then the quasiparticle
weight becomes $z_F=0.77$\cite{Mahan}. ]

In summary, we see that the spin correlation is reduced by a factor
of about two (from its maximum value one) as soon as we inject the two
electrons (entangled or not) into separate leads consisting of {\it
interacting} Fermi liquids in their ground state.
These findings are quite
encouraging in view of experimental investigations, as they
demonstrate that the spin correlations of a pair of electrons
in a Fermi
liquid will indeed be preserved in time (albeit with a reduced amplitude)
as long as we can neglect spin-dependent forces such as spin-orbit
interaction and spin flips induced by spin impurities or nuclear spins
etc. Given the high purity of present-day GaAs 2DEG's and the possibility
of suppressing the dephasing effects of nuclear spins by dynamical spin
polarization\cite{the8}, it looks promising to use mobile electrons in
nanostructures as a means for quantum communication.  Similar
investigations \cite{LBS} of such spin correlations are under way for
non-equilibrium transport situations, as well as for leads containing
impurities or consisting of superconducting or non-Fermi liquid
materials, etc.

In conclusion, we believe that various aspects of quantum
communication have a high chance of being realized in the
not-too-distant future. As we have seen, all that is needed is one
single quantum gate which is attached to leads and which can be used
as a source of entanglement for mobile qubits along the lines proposed
here.  Although the realization of such a device is still an
experimental challenge at present we are optimistic that it is within
technological reach.

\acknowledgments We would like to thank G. Burkard, C. Bennett,
R. Cleve, and E.  Sukhorukov for useful discussions.

\pagebreak

\begin{figure}[htbp]
\epsfxsize=8cm
\epsfbox{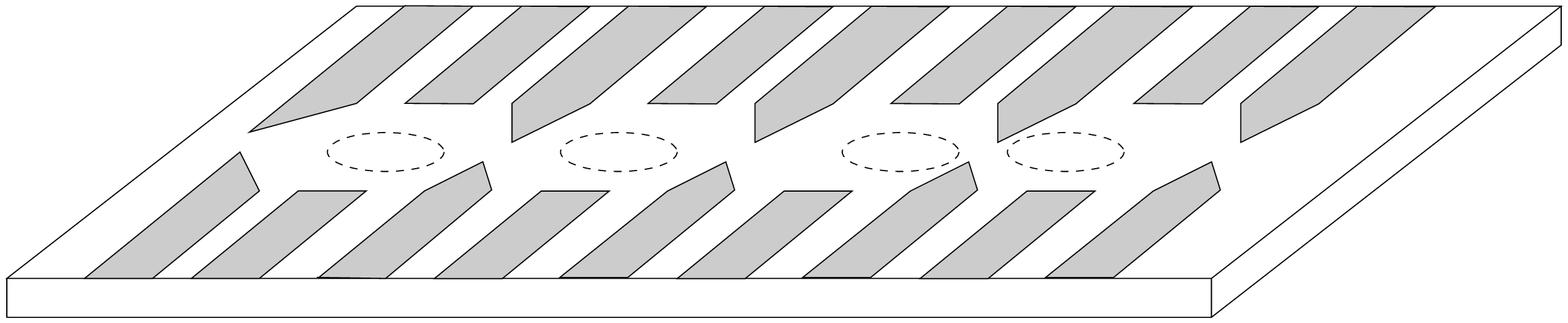}
\caption{A schematic of the quantum-dot array quantum computer.  Single
electrons are confined in a two-dimensional electron gas, and to dot
regions in between the electrodes.  Electrodes are shown shaded, dots
are shown as dashed circles.  The electrode potentials can be varied
so as to push pairs of electrons into contact (see the third and fourth
dots), which results in the execution of a two-bit quantum gate.  One-bit
gates are accomplished by the action of inhomogeneous magnetic fields
(or effective fields).  Readout is accomplished by tunneling the electrons
through a spin-selective barrier.  The magnetic elements of the device
are not shown.}
\label{fig1}
\end{figure}


\begin{references}
\bibitem{Brass} We are indebted to an excellent monograph,
G. Brassard, {\em Modern Cryptology, A Tutorial} (Springer-Verlag,
Lecture Notes in Computer Science, eds. G. Goos and J. Hartmanis,
Vol. 325, New York, 1988), for knowledge about much of the subject
matter of this section.  Our discussion of information processing with
quantum-mechanical tools is an update of Brassard's discussion in the
light of the many new advances in quantum information processing in
the last ten years.  See also A. J. Menezes, P. C. van Oorschot, and
S. A. Vanstone, {\em Handbook of Applied Cryptography} (CRC Press,
1996).
\bibitem{compl} C. H. Papadimitriou, {\em Computational Complexity},
(Addison-Wesley, 1994); M. Sipser, {\em Introduction to the Theory of
Computation} (PWS Pub. Co., 1997).
\bibitem{ccbook} E. Kushelevitz and N. Nisan, {\em Communication
Complexity} (Cambridge University Press, 1997).  The concept of
communication complexity was introduced in A. C.-C. Yao, ``Some
complexity questions related to distributive computing,'' {\em
Proc. of the 11th ACM Symp. on the Theory of Computing} (ACM Press,
1979), p. 209.
\bibitem{Abr} N. Abramson, {\em Information Theory and Coding}
(McGraw-Hill, New York, 1963).
\bibitem{sham} A. Shamir, ``How to share a secret,'' Comm. of the 
ACM, {\bf 22}, 612 (1979).
\bibitem{gamescl} J. von Neumann and O. Morgenstern, {\em Theory of
Games and Economic Behavior}, 3rd ed. (Princeton University Press,
Princeton, 1953).
\bibitem{PS} P. W. Shor, ``Polynomial time algorithms for prime
factorization and discrete logarithms on a quantum computer,'' SIAM
J. Comput. {\bf26}, 1484 (1997), and references therein.  
\bibitem{polys} R. Beals, H. Buhrman, R. Cleve, M. Mosca, and R. de
Wolf, ``Quantum lower bounds by polynomials,'' {\em Proc. of the 39th
Annual Symposium on the Foundations of Computer Science} (IEEE Press,
Los Alamitos, 1998), p. 352; quant-ph/9802049.
\bibitem{Ozhi} Y. Ozhigov, ``Quantum computer cannot speed up iterated
applications of a black box,'' quant-ph/9712051.
\bibitem{moreOzhi} Y. Ozhigov, ``Quantum computers speed up classical
with probability zero,'' quant-ph/9803064; E. Farhi, J. Goldstone,
S. Gutmann, and M. Sipser, ``A limit on the speed of quantum
computation in determining parity,'' Phys. Rev. Lett. {\bf 81}, 5442
(1998), quant-ph/9802045; ``A limit on the speed of quantum
computation for insertion into an ordered list,'' quant-ph/9812057;
``How many functions can be distinguished with {\em k} quantum
queries?'' quant-ph/9901012.
\bibitem{cleve98} H. Buhrman, R. Cleve, and A. Wigderson, ``Quantum
vs. Classical Communication and Computation,'' in {\em Proc. of the
30th Ann. ACM Symp. on the Theory of Computing} (ACM Press, 1998),
p. 63; eprint quant-ph/9802040.
\bibitem{Gro} L. K. Grover, ``Quantum mechanics helps in searching for
a needle in a haystack,'' Phys. Rev. Lett. {\bf 79}, 325 (1997).
\bibitem{CB} First done by R. Cleve and H. Buhrman, ``Substituting
quantum entanglement for communication,'' Phys. Rev. A {\bf56},
1201 (1997); quant-ph/9704026.
\bibitem{Ambai} A. Ambainis, L. Schulman, A. Ta-Shma, U. Vazirani, and
A. Wigderson, ``The quantum communication complexity of sampling,''
{\em Proc. of the 39th Annual Symposium on the Foundations of Computer
Science} (IEEE Press, Los Alamitos, 1998); see
http://www.icsi.berkeley.edu/\~\,amnon/Papers/qcc.ps.
\bibitem{Raz} R. Raz, ``Exponential separation of classical and
quantum communication complexity,'' {\em Proc. of the 31th Ann. ACM
Symp. on the Theory of Computing} (ACM Press, 1999), to be published.
\bibitem{bc} D. Mayers, ``Unconditionally secure quantum bit
commitment is impossible,'' Phys. Rev. Lett. {\bf 78}, 3414 (1997);
H.-K. Lo and H. F. Chau, ``Is quantum bit commitment really
possible?'' {\em ibid.} {\bf 78}, 3410 (1997); H.-K. Lo and
H. F. Chau, ``Why quantum bit commitment and ideal quantum coin
tossing are impossible,'' Physica D {\bf 120}, 177 (1998);
quant-ph/9711065.
\bibitem{Fuchs} C. A. Fuchs, ``Nonorthogonal quantum states maximize
classical information capacity,'' Phys. Rev. Lett. {\bf 79}, 1163
(1997) (see quant-ph/9703043); C. H. Bennett, C. A. Fuchs, and J. A.
Smolin, ``Entanglement-enhanced classical communication on a noisy
quantum channel,'' in {\em Quantum Communication, Computing, and
Measurement}, eds. O. Hirota, A. S. Holevo, and C. M. Caves (Plenum
Press, NY, 1997), p. 79; quant-ph/9611006.
\bibitem{superd} C. H. Bennett and S. J. Wiesner, ``Communication via
one- and two-particle operators on Einstein-Podolsky-Rosen states,''
Phys. Rev. Lett. {\bf 69}, 2881 (1992).
\bibitem{priv} C. H. Bennett, P. Shor, and J. A. Smolin, private
communication
\bibitem{qmoney} S. Wiesner, ``Conjugate coding,'' SIGACT News {\bf
15}, (1), 78 (1983).  [Manuscript prepared in 1970.]
\bibitem{Caves} C. M. Caves, K. S. Thorne, R. W. P. Drever,
V. D. Sandberg, and M. Zimmermann, ``On the measurement of a weak
classcial force coupled to a quantum-mechanical oscillator. I. Issues
of principle,'' Rev. Mod. Phys. {\bf 52}, 341 (1980).
\bibitem{BB84} C. H. Bennett and G. Brassard, ``Quantum Cryptography:
Public Key Distribution and Coin Tossing,'' in {\em Proceedings of the
IEEE International Conference on Computers, Systems and Signal
Processing, Bangalore, India} (IEEE, New York, 1984), p. 175.
\bibitem{BB84sec} D. Mayers, ``Unconditional security in quantum
cryptography,'' quant-ph/9802025.
\bibitem{QPAsec} H.-K. Lo and H. F. Chau, ``Quantum computers render
quantum key distribution unconditionally secure over arbitrarily long
distance,'' quant-ph/9803006.
\bibitem{Weg} M. N. Wegman and J. L. Carter, ``New hash functions and
their use in authentication and set equality,'' J. of Computer and
System Sciences {\bf22}, 265 (1981).
\bibitem{g8etal} C. H. Bennett, D. P. DiVincenzo, C. A. Fuchs, T. Mor,
E. Rains, P. W. Shor, J. A. Smolin, and W. K. Wootters, ``Quantum
nonlocality without entanglement,'' Phys. Rev. A {\bf59}, 1070 (1999);
quant-ph/9804053); M. Hillery, V. Buzek, and A. Berthiaume, ``Quantum
secret sharing,'' quant-ph/9806063; A. Karlsson, M. Koashi, and
N. Imoto, ``Quantum entanglement for secret sharing and secret
splitting,'' Phys. Rev. A {\bf 59}, 162 (1999).
\bibitem{games} D. A. Meyer, ``Quantum strategies,'' quant-ph/9804010;
J. Eisert, M. Wilkens, and M. Lewenstein, ``Quantum games and quantum
strategies,'' quant-ph/9806088; L. Goldenberg, L. Vaidman, and S.
Wiesner, ``Quantum gambling,'' quant-ph/9808001.
\bibitem{Preskill} J. Preskill, ``Reliable quantum computers,'' Proc.
R. Soc. London {\bf 454}, 385 (1998); quant-ph/9705031.
\bibitem{CGL} R. Cleve, D. Gottesman, and H.-K. Lo, ``How to share a
quantum secret,'' quant-ph/9901025.
\bibitem{BDSW} C. H. Bennett, G. Brassard, S. Popescu, B. Schumacher,
J. A. Smolin, and W. K. Wootters, ``Purification of noisy entanglement
and faithful teleportation via noisy channels,'' Phys. Rev. Lett. {\bf
76}, 722 (1996), eprint quant-ph/9511027; C. H. Bennett,
D. P. DiVincenzo, J. A. Smolin, and W. K. Wootters, ``Mixed state
entanglement and quantum error-correction,'' Phys. Rev. A {\bf54},
3824 (1996); e-print quant-ph/9604024.
\bibitem{bbcjpw} C. H. Bennett, G. Brassard, C. Crepeau, R. Jozsa,
A. Peres, W. K. Wootters, ``Teleporting an Unknown Quantum State via
Dual Classical and Einstein-Podolsky-Rosen Channels,'' Phys.
Rev. Lett. {\bf 70}, 1895 (1993).
\bibitem{LBS} D. Loss, G. Burkard, and E. Sukhorukov,
``Quantum communication with electrons", unpublished.
\bibitem{Atac} A. Imamoglu {\em et al.}, ``Quantum dot electron spin
manipulation using cavity QED,'' unpublished.
\bibitem{NIST} C. Monroe, D. M. Meekhof, B. E. King, W. M. Itano, and
D. J. Wineland, ``Demonstration of a fundamental quantum logic gate,''
Phys. Rev. Lett. {\bf75}, 4714 (1995).
\bibitem{NIST2} Q. A. Turchette, C. S. Wood, B. E. King, C. J. Myatt,
D. Liebfried, W. M. Itano, C. Monroe, and D. J. Wineland,
``Deterministic entanglement of two trapped ions,''
Phys. Rev. Lett. {\bf81}, 3631 (1998).
\bibitem{CIT} Q. A. Turchette, C. J. Hood, W. Lange, H. Mabuchi, and
H. J. Kimble, ``Measurement of conditional phase shifts for quantum
logic,'' Phys. Rev. Lett. {\bf75}, 4710 (1995).
\bibitem{QC} P. D. Townsend, J. G. Rarity, and P. R. Tapster,
Electronics Letters {\bf29} (14), 1291 (1993); R. J. Hughes,
D. M. Alde, P. Dyer, G. G. Luther, G. L. Morgan, and M. Schauer,
``Quantum Cryptography,'' Contemp. Phys. {\bf36}, 149 (1995);
A. Muller, T. Herzog, B. Huttner, W. Tittel, H. Zbinden, and N. Gisin,
```Plug and Play' systems for quantum cryptography,''
Appl. Phys. Lett. {\bf70}, 793 (1997).
\bibitem{SD} K. Mattle, H. Weinfurter, P. G. Kwiat, and
A. Zeilinger, ``Dense coding in experimental quantum communcation,''
Phys. Rev. Lett. {\bf76}, 4656 (1996).
\bibitem{rometp} D. Boschi, S. Branca, F. De Martini, L. Hardy, and
S. Popescu, ``Experimental Realization of Teleporting an Unknown Pure
Quantum State via Dual Classical and Einstein-Podolsky-Rosen
Channels,'' Phys. Rev. Lett. {\bf80}, 1121 (1998).
\bibitem{innsbrucktp} D. Bouwmeester, Jian Wei Pan, K. Mattle,
M. Eibl, H.  Weinfurter, and A. Zeilinger, ``Experimental Quantum
Teleportation,'' Nature {\bf390}, 575 (1997).  
\bibitem{cittp} L. Vaidman, Phys. Rev. A {\bf49} 1473 (1994);
A. Furusawa, J. L. Sorensen, S. L. Braunstein, C. A. Fuchs,
H. J. Kimble, and E. S.  Polzik, Science {\bf282}, 706 (1988).
\bibitem{Div95} D. P. DiVincenzo, ``Two-bit gates are universal for
quantum computation,'' Phys. Rev. A {\bf51}, 1015 (1995),
cond-mat/9407022.
\bibitem{NMR} N. Gershenfeld and I. Chuang, Science {\bf275}, 350
(1997); D. Cory, A. Fahmy, and T. Havel, Proc. Nat. Acad. Sci.
{\bf94} (5), 1634 (1997).
\bibitem{Ike} I. L. Chuang, L. Vandersypen, D. Leung, X. Zhou, and S.
Lloyd, Nature {\bf393}, 143 (1998); I. L. Chuang, N. Gershenfeld, and
M. G. Kubinec, Phys. Rev. Lett. {\bf80}, 3408 (1998).
\bibitem{Cory} D. G. Cory, M. D. Price, W. Maas, E. Knill, R. Lafamme,
W. H. Zurek, T. F. Havel, and S. S. Somaroo, ``Experimental quantum
error correction,'' Phys. Rev. Lett. {\bf81}, 2152 (1998).
\bibitem{nmrtp} M. A. Nielsen, E. Knill, and R. Laflamme, Nature
{\bf396}, 52 (1998).
\bibitem{the5} D. P. DiVincenzo, ``Topics in Quantum Computers,'' in
{\em Mesoscopic Electron Transport}, ed. L. Sohn, L. Kouwenhoven, and
G. Schoen (NATO Advanced Study Institute, Series E (Applied Sciences),
Vol. 345, Kluwer, 1997), p. 657, eprint (1996) cond-mat/9612126.
\bibitem{the6} D. Loss and D. P. DiVincenzo, ``Quantum computation
with quantum dots,'' Phys. Rev. A {\bf 57}, 120 (1998), eprint (1997)
cond-mat/9701055.
\bibitem{the7} D. P. DiVincenzo and D. Loss, ``Quantum information is
physical,'' Superlattices and Microstructures {\bf 23}, 419 (1998),
eprint (1997) cond-mat/9710259.
\bibitem{the8} G. Burkard, D. Loss, and D. P. DiVincenzo, ``Coupled
quantum dots as quantum gates,'' Phys. Rev. B, in press (Jan. 15,
1999); eprint cond-mat/9808026.
\bibitem{jacak} L. Jacak, P. Hawrylak, and A. W\'{o}js, {\em Quantum
Dots} (Springer, Berlin, 1997).
\bibitem{kouwenhoven} L.~P. Kouwenhoven, C.~M. Marcus, P.~L. McEuen,
S. Tarucha, R.~M. Westervelt, and N.~S. Wingreen, Proceedings of the
Advanced Study Institute on {\em Mesoscopic Electron Transport},
edited by L.~L. Sohn, L.~P. Kouwenhoven, G. Sch\"on (Kluwer, 1997).
\bibitem{ashoori} R.~C. Ashoori, Nature {\bf 379}, 413 (1996).
\bibitem{kotthaus} R.~J. Luyken, A. Lorke, M. Haslinger, B.~T. Miller,
M. Fricke, J.~P. Kotthaus, G. Medeiros-Ribiero, and P.~M. Petroff,
preprint.
\bibitem{tarucha} S. Tarucha, D.~G. Austing, T. Honda, R.~J. van der
Hage, and L.~P. Kouwenhoven, Phys.\ Rev.\ Lett.\ {\bf 77}, 3613
(1996); L.~P. Kouwenhoven, T.~H. Oosterkamp, M.~W.~S. Danoesastro,
M. Eto, D.~G. Austing, T. Honda, and S. Tarucha, Science {\bf 278},
1788 (1997).
\bibitem{waugh} F.~R. Waugh, M.~J. Berry, D.~J. Mar, R.~M. Westervelt,
K.~L. Chapman, and A.~C. Gossard, Phys.\ Rev.\ Lett.\ {\bf 75}, 705
(1995); C. Livermore, C.~H. Crouch, R.~M. Westervelt, K.~L. Chapman,
and A.~C. Gossard, Science {\bf 274}, 1332 (1996).
\bibitem{oosterkamp} T.~H. Oosterkamp, S.~F. Godijn, M.~J. Uilenreef,
Y.~V. Nazarov, N.~C. van der Vaart, and L.~P. Kouwenhoven, Phys.\
Rev.\ Lett.\ {\bf 80}, 4951 (1998).
\bibitem{Kik} J. M. Kikkawa, I. P. Smorchkova, N. Samarth, and D. D.
Awschalom, ``Room-temperature spin memory in two-dimensional electron
gases,'' Science {\bf 277}, 1284 (1997).
\bibitem{GAPA} J. A. Gupta, D. D. Awschalom, X. Peng, and A. P. Alivisatos,
``Spin coherence in semiconductor quantum dots,'' Phys. Rev. B, in
press.
\bibitem{crock} D. P. DiVincenzo, ``Two-bit gates are universal for
quantum computation,'' Phys. Rev. A {\bf 51}, 1015 (1995),
cond-mat/9407022.
\bibitem{blick} R.~H. Blick, D. Pfannkuche, R.~J. Haug, K. v.\
Klitzing, and K. Eberl, Phys.\ Rev.\ Lett.\ {\bf 80}, 4032 (1998).
\bibitem{Oosterkampmolecule} T.~H. Oosterkamp, J.~W. Janssen, L.~P.
Kouwenhoven, D.~G. Austing, T. Honda, S. Tarucha, cond-mat/9809142.
\bibitem{Barenco} A. Barenco, C. H. Bennett, R. Cleve,
D. P. DiVincenzo, N. Margolus, P. Shor, T. Sleator, J. A. Smolin, and
H. Weinfurter, ``Elementary gates for quantum computation,''
Phys. Rev. A {\bf 52}, 3457 (1995), quant-ph/9503016.
\bibitem{Prinz} G. A. Prinz, Science {\bf 282} 1660 (1998).
\bibitem{Kleiber}
M. Kleiber, F. Kuemmerlen, M. Loehndorf, A. Wadas, D. Weiss,
R. Wiesendanger ``Magnetization switching of submicrometer Co dots
induced by a magnetic force tip,'' Phys. Rev. B {\bf 58}, 5563 (1998).
\bibitem{Gott} D. Gottesman, private communication.
\bibitem{CZ} J. I. Cirac and P. Zoller, ``Quantum computations with
cold trapped ions,'' Phys. Rev. Lett.  {\bf 74}, 4091 (1995).
\bibitem{spinf} D. P. DiVincenzo, ``Quantum computing and single-spin
measurements using the spin-filter effect,'' J. Appl. Phys, in press,
cond-mat/9810295.
\bibitem{Kane} B. E. Kane, ``A Si-based nuclear-spin quantum computer,''
Nature {\bf 393}, 133 (1998).
\bibitem{Viola} L. Viola and S. Lloyd, ``Dynamical suppression of
decoherence
in two-state quantum systems,'' Phys. Rev. A {\bf 58}, 2733 (1998),
quant-ph/9803057.
\bibitem{Aspect} A. Aspect, J. Dalibard, and G. Roger,
Phys.\ Rev.\ Lett.\ {\bf 49}, 1804 (1982).
\bibitem{Gisin} W. Tittel, J. Brendel, H. Zbinden, and N. Gisin,
Phys.\ Rev.\ Lett.\ {\bf 81}, 3563 (1998).
\bibitem{Einstein}
A. Einstein, B. Podolsky, and N. Rosen,
Phys.\ Rev.{\bf 47}, 777 (1935).
\bibitem{Brown}
R.\ Hanbury Brown and R.\ Q.\ Twiss, Nature (London) {\bf 177}, 27
(1956).  M.\ B\"{u}ttiker, Phys.\ Rev.\ B{\bf 46}, 12485 (1992).  R.\
C.\ Liu, B.\ Odom, Y.\ Yamamoto, and S.\ Tarucha, Nature {\bf 391},
263 (1998).
\bibitem{Fetter} A.L. Fetter and J.D. Walecka, {\em Quantum Theory of
Many-Particle Systems} (New York, McGraw-Hill, 1971).
\bibitem{Quinn} G. F. Guiliani and J. J. Quinn, Phys. Rev. B {\bf 26},
4421 (1982).
\bibitem{Mahan} G. D. Mahan, {\em Many Particle Physics}, 2nd Ed. (Plenum,
New York, 1993).

\end{references}
\end{document}